\documentclass[trackchanges,twocolumn]{aastex701}
\usepackage{amsmath}
\usepackage{amssymb}
\usepackage{hyperref}
\usepackage{soul}
\usepackage{graphicx} 
\usepackage{subfigure}

\begin{document}

\title{Turbulent AGN coronae as the origin of diffuse neutrinos up to PeV energies}

\author[0009-0003-7748-3733]{Qi-Rui Yang}
\affiliation{School of Astronomy and Space Science, Nanjing University, Nanjing 210023, China; xywang@nju.edu.cn}
\affiliation{Key laboratory of Modern Astronomy and Astrophysics (Nanjing University), Ministry of Education, Nanjing 210023, China}
\email{qr.yang@smail.nju.edu.cn}

\author[0000-0003-1576-0961]{Ruo-Yu Liu}
\affiliation{School of Astronomy and Space Science, Nanjing University, Nanjing 210023, China; xywang@nju.edu.cn}
\affiliation{Key laboratory of Modern Astronomy and Astrophysics (Nanjing University), Ministry of Education, Nanjing 210023, China}
\affiliation{Tianfu Cosmic Ray Research Center, Chengdu 610000, Sichuan, China}
\email{ryliu@nju.edu.cn}

\author[0000-0002-5881-335X]{Xiang-Yu Wang}
\affiliation{School of Astronomy and Space Science, Nanjing University, Nanjing 210023, China; xywang@nju.edu.cn}
\affiliation{Key laboratory of Modern Astronomy and Astrophysics (Nanjing University), Ministry of Education, Nanjing 210023, China}
\affiliation{Tianfu Cosmic Ray Research Center, Chengdu 610000, Sichuan, China}
\email{xywang@nju.edu.cn}

\begin{abstract}
It has been shown that the turbulence acceleration in AGN coronae can account for 1-10 TeV neutrinos  from some AGNs, such as the Seyfert galaxy NGC 1068. Based on this, there are attempts to explain the diffuse neutrinos  observed by IceCube with the accumulated contribution from a population of AGNs, but it is found that  the maximum neutrino energy   is less than tens of TeV, and as a result, additional source classes are needed to explain the high-energy component above this energy.
Recently, motivated by the detection of  $>100$ TeV neutrinos   from the Seyfert galaxy NGC 7469, it was shown that the turbulence acceleration in the corona can explain $>$100 TeV neutrinos  given a larger magnetization parameter ($\sigma\sim 1$) in the corona, which leads to a larger maximum proton energy and a hard proton spectrum. In this paper, we extend this assumption to the population of AGNs and study whether  the  population of AGNs with a wide range of magnetization can explain the entire diffuse neutrino flux.
We find  that AGN coronae could account for the diffuse neutrinos up to  PeV energies if a significant fraction of AGNs  have  magnetizations as large as $\sigma\sim 1-10$. 
This conclusion is insensitive to the shape of the magnetization parameter distribution  as long as  the range of the magnetization parameter  is sufficiently wide and the distribution is flat towards high magnetization. 
Interestingly, this model can also  explain the peak of the diffuse neutrino spectrum at $\sim30$ TeV.

\end{abstract}

\section{Introduction}
In 2013, the IceCube Collaboration reported the first observation of an astrophysical neutrino
flux, with energies extending up to the PeV-scale \citep{IceCube2013Sci...342E...1I}.
The spectrum of the diffuse neutrinos is roughly consistent with a power-law
over the energy range $\sim$ 10 TeV–10 PeV, with a power-law $\Gamma$= 2.4-2.9 depending on the specific analysis considered \citep{Abbasi2021PhRvD.104b2002A,Abbasi2022ApJ...928...50A}.  The latest results extend the spectrum down to  $\sim 1$ TeV and reveal evidence for a low-energy break near $\sim 30\,{\rm TeV}$ \citep{IceCube2025arXiv250722233A,Basu2025arXiv250706002B}, indicating that the diffuse neutrino spectrum
deviates from a simple power law with  a significance exceeding $4 \sigma$. The origin of this  diffuse neutrino emission remains unknown.

A complementary source of information is the identification of specific candidates for point sources of neutrinos. The IceCube collaboration \citep{2022Sci...378..538I} reported an excess of neutrino events associated with NGC 1068, a nearby type-2 Seyfert galaxy,
with a significance of $ 4.2\sigma$.  The reported neutrino flux is significantly higher than the GeV gamma-ray flux of the
galaxy \citep{Ackermann2012ApJ...750....3A}. It has hence been suggested that the  opaque AGN cores  can be the high-energy neutrino sources, where
dense radiation attenuates gamma-rays while providing abundant targets for neutrino production \citep{Murase2020ApJ...902..108M,Murase2022ApJ...941L..17M,Inoue2020ApJ...891L..33I,Kheirandish2021ApJ...922...45K,Halzen2022arXiv220200694H,Kurahashi2022ARNPS..72..365K,Eichmann2022ApJ...939...43E,Halzen2023arXiv230507086H}. Specifically, magnetized coronae are suggested to be promising
accelerators of cosmic-ray protons that  produce neutrinos mainly
through interactions with coronal X-rays \citep{Stecker1991PhRvL..66.2697S,Inoue2019ApJ...880...40I,Murase2022ApJ...941L..17M}.

The neutrino emission of NGC 1068 suggests a soft spectrum in the 1.5–15 TeV. This spectrum suggests that the parent proton spectrum is intrinsically soft in the corresponding region of proton energies around 30-300 TeV. These features can be accommodated in a proton-electron corona ($n_p=n_e$) with plasma beta $\beta = 8\pi n_p k T_p / B^2 \sim 1$ \citep{Murase2020PhRvL.125a1101M,2025A&A...697A.124L,Inoue2020ApJ...891L..33I,Eichmann2022ApJ...939...43E,Inoue2022arXiv220702097I,Yuan2025arXiv250808233Y}, which is equivalent to the magnetization
$\sigma =B^2/(4\pi n_p m_p c^2)=  2k T_p/(\beta m_p c^2) \sim 0.1 $, where $k$ is Boltzmann constant and $T_p$ is proton temperature, $B$ is the magnetic field strength, $n_p$ and $n_e$ are, respectively, the number densities of thermal protons and  electrons, which dominate the total number densities of protons and electrons, respectively.
Motivated by the detection of TeV neutrinos from NGC 1068, it was shown that the AGN population could account for the diffuse neutrino signal observed by IceCube in the ∼1–100 TeV energy range \citep{Murase2020PhRvL.125a1101M,Fiorillo2025ApJ...989..215F}, but is insufficient to explain higher energy neutrinos. 
An additional source classes, such as blazars \citep{Padovani2015MNRAS.452.1877P,Padovani2024A&A...684L..21P, Fiorillo2025ApJ...989..215F,Karavola2026arXiv260101533K} or accretion flows in low-luminosity AGNs \citep{2021NatCo..12.5615K,2015ApJ...806..159K}, have been invoked to explain the high-energy component above 100 TeV \citep{Padovani2015MNRAS.452.1877P,Padovani2024A&A...684L..21P, Fiorillo2025ApJ...989..215F,Karavola2026arXiv260101533K}.

Recently,  two 100 - 200 TeV neutrinos,  issued as two
neutrino alerts IC220424A and IC230416A, are found to be coincident with the Seyfert galaxy NGC 7469 \citep{Sommani2025ApJ...981..103S}, suggesting it
as a potential emitter of neutrino above 100 TeV. The likelihood fit of the IceCube data for this source returned a very hard spectral index of $\sim 1.9$ \citep{IceCube2025arXiv251013403A}.  We propose that cosmic rays with a larger maximum  energy accelerated in the turbulent corona,  achieved through a larger magnetization parameter ($\sigma\sim 1$)\citep{Yang2025ApJ...995..166Y}, can explain the neutrinos  from NGC 7469.  

In this paper, we assume that a fraction of X-ray AGNs has large magnetization parameter and pair-dominated composition in the corona, and investigate the contribution of the entire population of AGNs to the diffuse  neutrinos within the framework of the turbulent coronal model. For a single AGN, protons are accelerated by the turbulence in a magnetized corona, with their acceleration properties primarily determined by two independent parameters: the X-ray luminosity $L_X$ and the coronal magnetization $\sigma$. High-energy neutrinos are produced via $p\gamma$ interactions between relativistic protons and soft photon fields originating from the accretion disk and the  corona. We compute the resulting diffuse neutrino spectrum by convolving the single-source neutrino emission with the AGN X-ray luminosity function and an assumed distribution of the coronal magnetization across the AGN population.

In Section \ref{Sec:2}, we describe the physical properties of AGN coronae, including the coronal magnetization, plasma composition, and proton acceleration in turbulent coronae. In Section \ref{Sec:3}, we study the diffuse neutrino flux contributed by the AGN population. In Section \ref{Sec:4}, we present the conclusions and discussions.

\section{Properties of Corona and proton acceleration}
\label{Sec:2}
The hard X-ray emission from AGNs is commonly
attributed to a spatially compact region near the black
hole, referred to as the “corona”. We consider a generic setup in which a supermassive black hole with a mass of $M_{\rm BH}$ and gravitational radius $R_{\rm g}$ is embedded in the luminous radiation of accretion disk and corona. For simplicity, the corona that encompasses the inner accretion disk is assumed to be quasi-spherical, with a characteristic length of $R_{\rm c} = \mathcal{R}R_{\rm g}\simeq 10 (\mathcal{R}/10) R_{\rm g}$. To model the cumulative diffuse neutrino emission from the AGN population, we characterize the corona with two parameters: the X-ray luminosity $L_X$ in the 2-10 keV band, describing the coronal radiation field in soft X-rays and serving as a proxy of the proton luminosity $L_p$, and the magnetization $\sigma$, describing the plasma properties.

\subsection{The properties of coronal plasma}
In the standard disk-corona scenario, the observed X-rays are attributed to thermal Comptonlization of disk photons by high-temperature ($\sim 10^9\,{\rm K}$) electrons, and these electrons are presumably heated in the coronal region. Despite the rich observational evidence for the existence of hot coronae, there is no general consensus regarding the mechanism by which such coronae are formed and heated to such a high temperature. Many models
\citep{Haardt1991ApJ...380L..51H, Svensson1994ApJ...436..599S}
simply assume that some fraction of the energy liberated by accretion is dissipated in optically thin regions. Magnetic reconnection is often invoked as the energy dissipation mechanism in the corona
\citep{DiMatteo1997MNRAS.291L..23D,Merloni2001MNRAS.321..549M,Liu2002ApJ...572L.173L,Sironi2020ApJ...899...52S}. 
Recently, by utilizing ALMA and radio observations of  nearby radio-quiet Seyferts IC 4329A and
NGC 985, \cite{Inoue2018ApJ...869..114I} constructed the coronal radio synchrotron emission spectra  up to 250 GHz, and found that the magnetic
field strength is weaker than the prediction from the
magnetically heated accretion corona scenario. 
If this analysis is correct, for these AGNs, some other mechanisms  than the magnetic reconnection are needed for the heating of the corona. Possible mechanisms include the dissipation of turbulence driven by magneto-rotational instability \citep{Jiang2014ApJ...784..169J} and the evaporation of the inner accretion disk \citep{Meyer2000A&A...361..175M,2000A&A...360.1170R}. 

The magnetization parameter is extensively employed in both particle-in-cell (PIC) and magnetohydrodynamic (MHD) simulations to describe the dynamical behaviors and particle-acceleration characteristics of plasma within the coronae. 
The magnetization of the corona is defined  by
\begin{equation}
\sigma = \frac{B^{2}}{4\pi (n_p m_p+n_e m_e) c^{2}},
\label{Eq:Magnetization}
\end{equation}
which quantifies the ratio of magnetic energy density to the rest-mass energy density of thermal protons and electrons. If the plasma mass density is dominated by protons, i.e. $n_p m_p > n_e m_e$, the magnetization can be approximated as $\sigma \simeq B^2/(4\pi n_p m_p c^2)$.  In terms of  Alfv\'enic velocity $v_{\rm A}$, $\sigma = \beta_{\rm A}^2/(1-\beta_{\rm A}^2)$, with $\beta_A \equiv v_{\rm A}/c$. Given the uncertainties in the heating mechanism of the coronae, we assume that  the magnetization is  a free parameter independent of the X-ray luminosity and its distribution across the population of AGNs is unknown from first principles.

The composition of black-hole coronae remains largely uncertain. In modeling of the neutrinos from NGC 1068, a proton (ion)–electron composition ($n_e= n_p$) are usually assumed for the corona \citep{Murase2020PhRvL.125a1101M,Fiorillo2024ApJ...974...75F,2025A&A...697A.124L}. Nevertheless, as shown by \cite{Yang2025ApJ...995..166Y}, a corona characterized by a high magnetization and a higher density of electrons than protons ($n_e> n_p$) is necessary for producing neutrinos  with energies  over $100\,\rm{TeV}$ from NGC 7469. 

Motivated by the difference in the two AGNs, we consider a normal proton-electron composition  ($n_e= n_p$) in the low magnetization regime and allow a pair-dominated composition ($n_e\ge n_p$) in the high magnetization regime. 

Regardless of the coronal composition, the total electron–positron number density, $n_e$, is self-regulated such that the Thomson optical depth remains at $\tau_{\rm T}\sim 1$ \citep{Fabian2015MNRAS.451.4375F}, resulting in 
\begin{equation}
{n_{e}}\sim \tau_{\rm T}/\sigma_T R_{\rm c} \sim 10^{11}\,{\rm cm}^{-3}\left(\frac{10}{\mathcal{R}}\right)\left(\frac{10^7 M_{\odot}}{M_{\rm BH}}\right). 
\label{Eq:electron number density}
\end{equation}

Since we always have $n_p\leq n_e$, the maximally allowable magnetic field strength inferred from a given magnetization parameter $\sigma$  is given by (using Eq.\ref{Eq:Magnetization})
\begin{equation}
\begin{aligned}
    B_{\rm \sigma, max} &\simeq \sqrt{4 \pi \sigma n_e m_p c^2} \\
    &\simeq 3\times10^3 \,{\rm G}
    \left(\frac{10}{\mathcal{R}}\right)^{1/2}
    \left(\frac{10^7 M_{\odot}}{M_{\rm BH}}\right)^{1/2}
    \left(\frac{\sigma}{0.01}\right)^{1/2},
    \label{Eq:Bsig}
\end{aligned}
\end{equation}
in the regime of $n_p m_p > n_e m_e$  that applies to our case. 

On the other hand, the magnetic field  strength in the corona is  related with the X-ray luminosity of AGNs if the corona heating is dominated by the magnetic energy dissipation.
The energy dissipation through processes such as  reconnection  could represent an important means to power the X-ray emission \citep{Beloborodov2017ApJ...850..141B,2024PhRvL.132h5202G,Fiorillo2024ApJ...974...75F}. This implies $L_X \simeq \eta_X u_B  V_c/t_{\rm diss}$, where $\eta_X$ is the equipartition factor,
$V_c$ denotes the coronal volume, $u_B = B^2/8\pi$ is the magnetic energy density, $t_{\rm diss} \simeq \ell_c/(0.1v_{\rm A})\simeq \ell_c/(0.1c\sqrt{\sigma/(1+\sigma)})$ is the mean dissipation time of magnetic energy, also corresponding to a few eddy turn-over times,  $v_{\rm A}$ is the Alfv\'enic velocity, and $\ell_c\simeq R_{\rm g}$ is the outer scale of the turbulence. Requiring that the magnetic energy dissipation can account for the observed X-ray luminosity leads to an equipartition magnetic field strength,
\begin{equation}
\begin{aligned}
    &B_{\rm eq}\simeq 
    \sqrt{\frac{8\pi L_X }{\eta_X V_c\times 0.1c\sqrt{\sigma/(1+\sigma)}/\ell_c}}
    \simeq 6\times 10^3 \,{\rm G}\\
    &\left(\frac{1}{\sigma}+1\right)^{1/4}
    \left(\frac{10}{\mathcal{R}} \right)^{3/2}\left(\frac{0.1}{\eta_X}\right)^{1/2}
    \left(\frac{10^7M_{\odot}}{ M_{\rm BH}} \right)
    \left(\frac{L_X}{10^{42}\,{\rm erg\,s^{-1}}}\right)^{1/2},
  \label{Eq:Beq}
\end{aligned}
\end{equation}
which represents the maximum magnetic field strength allowed by the X-ray energy budget.
In the case $B_{\rm eq}> B_{\rm \sigma,max} $, the magnetic energy dissipation is insufficient to power the X-ray emission of the corona. As this case applies usually to the low magnetization case, we have a proton-electron composition typically, so the magnetic field strength of the corona is  roughly $B=B_{\rm \sigma,max}$. 
On the other hand, if $B_{\rm eq}\leq B_{\rm \sigma,max} $, the magnetic energy dissipation power becomes comparable to the observed X-ray luminosity, so the magnetic dissipation plays the dominant role in the coronal heating. 
In this high-$\sigma$ regime, the plasma is  expected to become pair-dominated ($n_p < n_e$), allowing the corona to remain highly magnetized. In this case, the magnetic field strength of the corona can be inferred from  the equipartition value (i.e., $B=B_ {\rm eq}$). 

In the following calculation of the diffuse neutrino flux from the population of AGNs, we treat the magnetization parameter $\sigma$ as an input free parameter. Accordingly, the magnetic field strength across the entire parameter space is taken to be 
\begin{equation}
    B \simeq \min (B_{\rm \sigma,max}, B_{\rm eq}).
    \label{Eq:B}
\end{equation}
To consistently account for the transition between proton–electron coronae and pair-dominated coronae, we parameterize the proton number density as
$n_p \simeq \min\!\left({B_{\rm eq}^{2}}/{4\pi \sigma m_p c^{2}},\, n_e \right)$.

\subsection{Maximum proton energy}
Particle acceleration in magnetically dominated turbulence is primarily regulated by the amplitude of magnetic fluctuations on the outer scale, $\ell_c \simeq R_{\rm g}$. The strength of the turbulent magnetic field is characterized by its root-mean-squared value $\delta B$. We suggest $\delta B$ is comparable to the mean coronal magnetic field $B$. This regime of large-amplitude, semi-relativistic turbulence has recently gained insight through large-scale particle-in-cell (PIC) numerical experiments~\citep[e.g.][]{Zhdankin2018ApJ...867L..18Z,ComissoSironi2019ApJ...886..122C,Wong2020ApJ...893L...7W,Bresci+22,2023ApJ...944..122M,2025MNRAS.543.1842W,2025arXiv250604212D}, and analytical developments~\citep[e.g.][]{2021PhRvD.104f3020L,2022PhRvL.129u5101L}. The mean energy diffusion coefficient measured in the above simulations is $D_{\gamma\gamma}\approx 0.1\, \gamma^2\,\sigma c/\ell_{\rm c}$. We can therefore write the turbulence acceleration timescale as 
\begin{equation}
    t_{\rm tur} = \frac{1}{4}\frac{\gamma^2}{D_{\gamma\gamma}} \simeq 2.5\,\sigma^{-1}\frac{\ell_c}{c}\simeq 100\, {\rm s} \,\sigma^{-1} \frac{M_{\rm BH}}{10^7M_{\odot}}.
\label{Eq: Turbulence}
\end{equation}

In AGN coronae, we account for all relevant proton cooling processes, including $pp$, $p\gamma$, Bethe–Heitler, and proton synchrotron radiation. In addition, we consider particles escape through both diffusive and in-fall processes. Note that the dominant cooling channels may vary among AGNs due to difference in their coronal properties. To reduce the number of free parameters, we adopt the empirical relation between the SMBH mass and the X-ray luminosity, as proposed by \citet{Mayers2018arXiv180306891M},
\begin{equation}
    M_{\rm BH} = 3\times10^7 M_{\odot}(L_X/2\times10^{43}\,{\rm erg\,s^{-1}})^{0.58}.
    \label{Eq:M_BH}
\end{equation}

\begin{figure}
    \centering
    \includegraphics[width=0.5\textwidth]{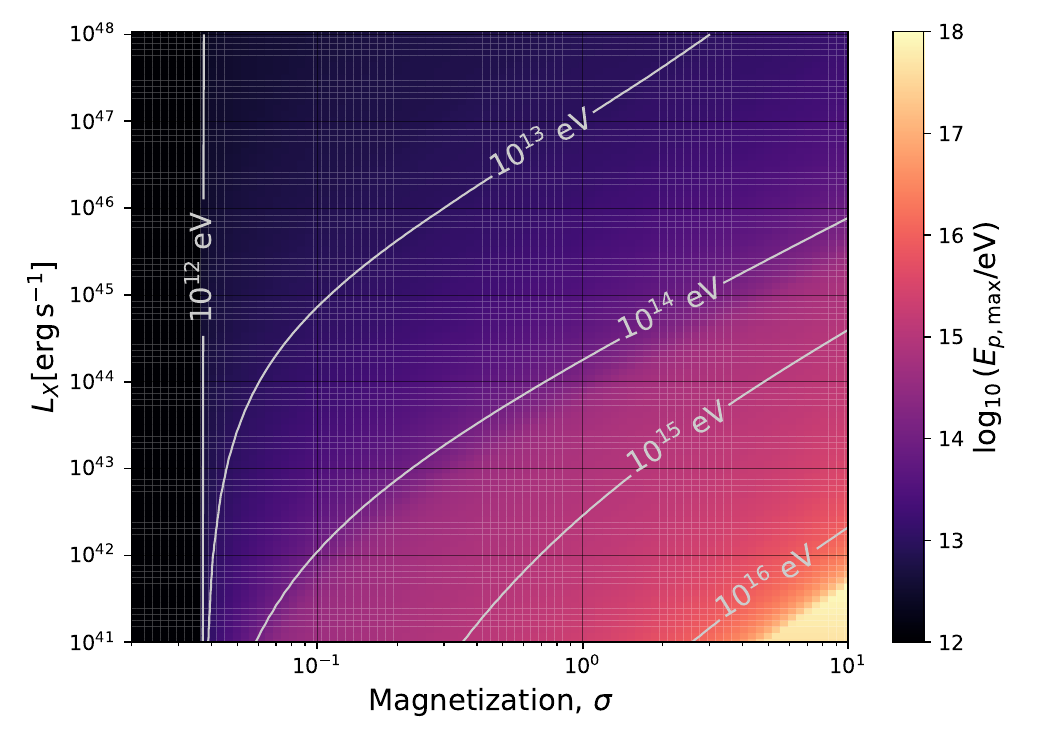}
    \caption{Color map of the maximum energy of protons as a function of X-ray luminosity ($L_X$) and magnetization ($\sigma$).   The left dark region represents the parameter space that protons can not be efficiently accelerated (i.e. $t_{\rm tur} \gtrsim t_{\rm loss}$ and the energy loss is dominated by $pp$ and/or in-fall). The bright region at the bottom-right corner represents the maximum proton energy constrained by the Hillas condition.}
    \label{Fig:Ep_max}
\end{figure}

Accordingly, we determine the maximum proton energy by equating the total energy-loss timescale $t_{\rm loss}$ (see Appendix for details) with the turbulent acceleration timescale $t_{\rm tur}$. The Hillas criterion is also considered as an additional constraint to prevent unrealistically high maximum energies within the outer scale of the turbulence, given by
\begin{equation}
    E_{p,{\rm Hillas}} = eB\ell_c
    \label{Eq:Hillas},
\end{equation}
where $e$ is the elementary charge in Gaussian units.
The  maximum energy of protons  as a function of $L_X$ and $\sigma$ is plotted in the color map Fig.\ref{Fig:Ep_max}. The color illustrate the magnitude of the maximum energy that protons can be accelerated. Since a lower X-ray luminosity implies a shorter acceleration timescale due to a decreased SMBH mass and coronal size (Eq.\eqref{Eq:M_BH}), the acceleration becomes more efficient. Accordingly, the bottom-right region in Fig. \ref{Fig:Ep_max} shows that highly magnetized coronae with lower X-ray luminosities can accelerate protons to higher energies, with the maximum proton energy constrained by the Hillas condition.

The dark region with $\sigma\lesssim 0.04$ on the very left side of Fig. \ref{Fig:Ep_max} corresponds to the regime of very inefficient acceleration with a maximum proton energy below 1 TeV, where the turbulent acceleration timescale satisfies $t_{\rm tur} \gtrsim t_{\rm loss}$. In this parameter space, the dominant energy-loss processes are $pp$ and/or escape due to in-fall.

\subsection{Proton spectrum}
Observations of the diffuse astrophysical neutrino flux indicate that the spectrum extends to $\sim 10\,{\rm PeV}$ as a broken power-law with a post-break spectral index of $\Gamma \sim 2.7$ and broken energy of $\sim 30 \,{\rm TeV}$ \citep{IceCube2025arXiv250722233A,Basu2025arXiv250706002B}. This suggests that sources whose neutrino spectra exhibit a spectral index softer than $\Gamma \sim 2.7$ below $\sim 30 \,{\rm TeV}$,  as detected from individual sources such as NGC 1068 or NGC 4151, are unlikely to dominate the high-energy diffuse neutrino background over 30 TeV. In contrast, sources such as NGC 7469, which exhibit neutrino emission extending to much higher energies with a hard spectra, may constitute the dominant contributors to the high-energy component of the diffuse neutrino flux. In our model, a key parameter in distinguishing these various classes of sources is the magnetization, as it may influence both the resulting spectral index and the highest achievable proton energy.

Turbulent acceleration is known to shape
hard momentum ($p$) distributions in the time-asymptotic
limit, of the form $dn/dp\propto p^{-\Gamma}$ with $\Gamma\sim1$. The turbulence may  undergo nonlinear feedback when the energy rate at which high-energy protons draw energy from the cascade becomes commensurable with the rate at which energy is injected into the turbulent cascade \citep{Lemoine2024PhRvD.109f3006L}.  In the regime of strong damping, the nonthermal particle energy spectrum takes on a generic broken power-law shape, with a hard slope at low energies
($\Gamma\le 1$) and flat segment at high energies ($\Gamma\sim 2$)\citep{Lemoine2024PhRvD.109f3006L}.

In practice, the feedback sets in when the proton energy density satisfy
\begin{equation}
    u_p \gtrsim \beta_A^{1-q} u_B \simeq  \left({\frac{\sigma}{\sigma+1}}\right)^{-2/3}u_B,
    \label{Eq:feedback}
\end{equation}
where $q = 5/3$ represents Kolmogorov turbulence and $v_{\rm A}\simeq \sqrt{\sigma/(\sigma+1)}c$ is the Alfv\'enic speed.

The break energy for the broken power-law spectrum is given by
\begin{equation}
\begin{aligned}
 E_{p,{\rm br}}&\simeq 10 E_0\left(\frac{v_A}{c}\right)^{-1} x_{\rm nth}^{-1}\sigma\\ 
 &\simeq 10\, {\rm GeV}\,\sigma \left(\frac{E_0}{m_pc^2}\right)\left(\frac{v_A}{c}\right)^{-1}\left(\frac{0.1}{x_{\rm nth}}\right),
\end{aligned}
\end{equation}
where $E_0 \sim m_p c^2$ denotes the energy at injection in microscopic reconnecting current sheets and $ x_{\rm nth}$ is the fraction of the thermal plasma converted into non-thermal particles.

{In turbulent coronae, high-energy protons are assumed to be energized by turbulent magnetic fields. Accordingly, we can parameterize the proton power as a fraction $\eta_p$ of the magnetic luminosity, $L_p = \eta_p L_B$.
The magnetic luminosity is determined by $L_B = (B^2/8\pi) V_c/ t_{\rm diss}$ with the magnetic field given by Eq.\eqref{Eq:B}.}
The proton energy density, $u_p$,  can be approximately obtained by assuming that the energy injected in high-energy protons is dissipated within a typical energy loss timescale $t_{\rm loss}$, i.e., $L_p \simeq u_p\times V_c/t_{\rm loss}$. Hence the ratio of $u_p$ and $u_B$ is given by
\begin{equation}
    \frac{u_p}{u_B} = \frac{L_p\times t_{\rm loss}}{V_c u_B} \simeq \eta_p \frac{t_{\rm fall}}{t_{\rm diss}}
    =\eta_p \frac{R_{\rm c}/V_r}{ \ell_c/(0.1v_{\rm A})}.
    \label{Eq:u_p/u_B}
\end{equation}
Here  $t_{\rm loss}$ is the timescale for total energy loss associated with the medium energy of the broken power-law spectrum characterized by $\sim E_{p,{\rm br}}$, where the energy loss is dominated by the in-fall  time $t_{\rm fall} \simeq R_{\rm c} / V_r$ ($V_r = \alpha V_k /2$ is radial velocity, $\alpha$ is the viscous parameter and $V_k = \sqrt{GM_{\rm BH}/R_{\rm c}}$ is the Keplerian velocity).  Then, combining Eq.\eqref{Eq:u_p/u_B} and Eq.\eqref{Eq:feedback}, we can obtain the feedback criterion
\begin{equation}
    \eta_p \gtrsim \frac{5\alpha}{\mathcal{R}^{3/2}}\left( \frac{\sigma}{\sigma+1} \right)^{-5/6} \sim  0.1 \sigma^{-5/6} \left( \frac{\alpha}{0.3} \right)\left( \frac{10}{\mathcal{R}} \right)^{-3/2}.
    \label{Eq:feedback criterion}
\end{equation}
Assuming a benchmark value of $\eta_p\sim 0.1$ for the AGN population, Eq.\eqref{Eq:feedback criterion} indicates that coronal turbulence commences feedback  when $\sigma$ exceeds unity for typical parameter values of $\alpha$ and $\mathcal{R}$.

\section{Diffusive neutrino emission from a population of AGN coronae}
\setcounter{footnote}{0}
\label{Sec:3}
In this section, we extend the above relations to the entire AGN population, constraining the diffuse neutrino fluxes of AGNs.  

The neutrino luminosity radiated from an AGN can be calculated as a function of $L_X$ and $\sigma$, given by
\begin{equation}
    L_{\nu}(L_X,\,\sigma)\simeq\frac{3}{8}f_{p\gamma}(L_X,\,\sigma) L_p(L_X),
    \label{Eq:single source}
\end{equation}
where $f_{p\gamma} \simeq t_{p\gamma}^{-1}/ t_{\rm loss}^{-1}$ is the  efficiency of neutrino production through the $p\gamma$ process. Here $t_{p\gamma}$ is the timescale of $p\gamma$ process defined by Eq.\eqref{Eq:timescale pgamma calculated} and $t_{\rm loss}$ is the timescale of total energy loss given by Eq.\eqref{Eq:loss}. In the energy range of interest ($E_{\nu}\gtrsim  10\, {\rm TeV}$), $pp$ interactions are negligible for both particle cooling and neutrino emission. Therefore, we only consider the contribution of $p\gamma$ interactions to neutrino production.

In our calculation, we adopt $\eta_p \sim 0.1$ as a benchmark value.
For comparison, using the neutrino luminosity of NGC 1068 $L_{\nu,\,{\rm NGC\,1068}} \simeq 2.9\times 10^{42}\,{\rm erg\,s^{-1}} \simeq \frac{3}{8}f_{p\gamma,\,pp}(\eta_p/\eta_X )L_{X,\,{\rm NGC\,1068}}$ \citep{2022Sci...378..538I}, and together with the observed X-ray luminosity $L_{X,\,{\rm NGC\,1068}}\simeq 7\times10^{43}\,{\rm erg\,s^{-1}}$ \citep{Marinucci2016MNRAS.456L..94M}, we can deduce the efficiency for NGC 1068, i.e., $\eta_p \simeq 0.1 \, (L_{\nu}/2.9\times10^{42}\,{\rm erg\,s^{-1}})(7\times10^{43}\,{\rm erg\,s^{-1}}/L_X)(0.1/f_{p\gamma,\,pp})(\eta_X/0.1)$.
We assume that sources capable of producing astrophysical neutrinos are as efficient as NGC 1068 in energizing relativistic protons with an efficiency of $\eta_p \sim 0.1 $.

We then calculate the diffuse neutrino flux from the AGN population.
For the distribution of X-ray luminosity across the AGN population, 
we adopt the prescription for the X-ray luminosity function (XLF) of \cite{Ueda2014ApJ...786..104U}, introducing $d\Phi/d {\rm log_{10}} L_X$ as the number of sources per co-moving volume per decade of luminosity. For the distribution on the magnetization parameter, we assign a random value $\sigma$ to each source, sampled from a log-normal probability distribution, namely,
\begin{equation}
    P(\sigma) \simeq \frac{1}{\sqrt{2\pi}\sigma\,s\, {\rm ln}10}\exp\!\left[-\frac{(\log_{10} \sigma-\mu)^2}{2s^2}\right], 
    \label{Eq:log-normal}
\end{equation}
where $\mu\simeq {\rm log_{10}}(0.1)=-1$, denoting that the log-normal distribution is centered at $\sigma = 0.1$, (0.1 is considered to be a fiducial value in AGN coronae), and $s \simeq 1$ is the broadening of the log-normal distribution. In this work, we restrict the magnetization to be in the range $0.04 \leq \sigma \leq 10$. The lower limit $\sigma \simeq 0.04$ corresponds to the minimum magnetization required for efficient non-thermal proton acceleration. The upper limit $\sigma \simeq 10$ is adopted from the simulation result in \cite{Liska2022ApJ...935L...1L}\footnote{\citet{Liska2022ApJ...935L...1L} presented a  GRMHD simulations of luminous sub-Eddington accretion disks. They find that the ratio between the plasma pressure to magnetic pressure  is $p_i/p_b \sim 0.01$ at $R=10 R_g$ in the case of purely poloidal magnetic
field, implying a maximum magnetization of $\sigma \sim 10$.}. We also normalize the distribution function over this range.

\begin{figure}
    \centering
    \includegraphics[width=0.5\textwidth]{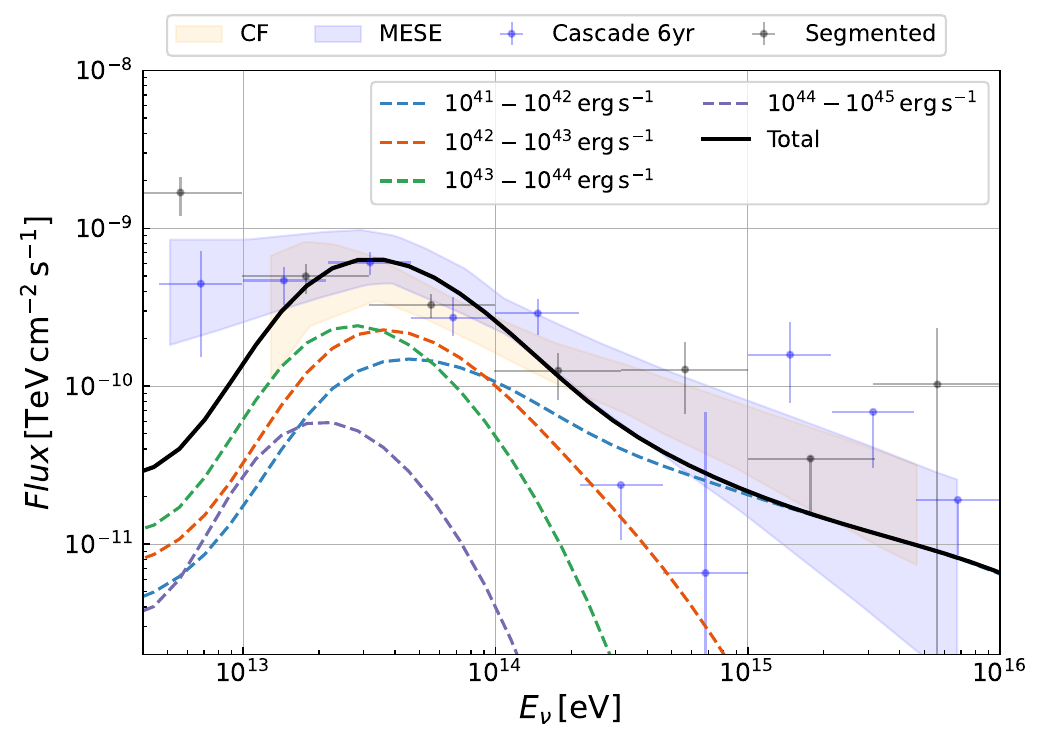}
    \caption{Diffuse all-flavor neutrino fluxes from the AGN coronae assuming that the magnetization distribution follows the log-normal distribution with parameters $\mu = -1$ and $s = 1$. The colored dashed curves denote the contributions from AGNs in the different ranges of the X-ray luminosity, while the black solid curve represents the sum of them. 
    The blue and black data points represent IceCube observations, while the orange and blue shaded regions show the broken power-law fit reported by IceCube \citep{IceCube2025arXiv250722233A}. In the calculations, we adopt $\mathcal{R} = 10$, $\eta_p \simeq 0.1$, $\eta_X \simeq 0.1$, and $\zeta \simeq 0.1$.} 
    \label{Fig:Diffuse1}
\end{figure}

By performing a convolution of the magnetization distribution given in Eq.\eqref{Eq:log-normal}, the XLF characterized by \cite{Ueda2014ApJ...786..104U}, and the neutrino flux from an individual source as defined by Eq.\eqref{Eq:single source}, the diffuse astrophysical neutrino flux can be obtained from
\begin{equation}
\begin{aligned}
    F_{\nu, {\rm diff}} = & 
    \zeta\frac{c}{H_0}\int^{z_{\rm max}}_{0}\frac{dz}{E(z)}\int^{10^{45}}_{10^{41}}\int^{\sigma_{\rm max}}_{\sigma_{\rm min}}L_{\nu} [E(z+1),L_X,\sigma] \\
    &\times P(\sigma)\frac{d\Phi}{dL_X}dL_X d\sigma,
    \label{Eq:Diffuse}
\end{aligned}
\end{equation}
where $\zeta$ represents a suppression factor of the neutrino flux, which could result from a duty cycle of neutrino production in  AGNs or only a fraction ($\zeta$) of AGNs contributing to the diffuse neutrino flux.  \cite{2025ApJ...988..141A} showed that the absence of a significant signal in the stacking analysis also indicate a suppression factor $\zeta$, which supports our assumption.
Fig. \ref{Fig:Diffuse1} illustrates the diffuse neutrino flux in comparison with the IceCube data. The colored dashed curves represent the contribution from AGNs within a certain X-ray luminosity range, while the black curve represents the sum of the contribution from all AGNs. We restrict our analysis to the luminosity range $10^{41}\,{\rm erg\,s^{-1}} \leq L_X \leq 10^{45}\,{\rm erg\,s^{-1}}$, since AGNs outside this X-ray luminosity range have a negligible population and thus provide an insignificant contribution to the diffuse flux. The blue and orange shades represents the  fitted neutrino spectrum by using two different analysis methods, namely Medium Energy Starting Events (MESE) and Combined Fit (CF) \citep{IceCube2025arXiv250722233A}. 

We find that the theoretically calculated diffuse neutrino flux significantly overshoots the IceCube measurements by up to an order of magnitude if $\zeta=1$, a contradiction that has also been  noted in previous studies \citep{2026PhRvD.113b3019S, Fiorillo2025ApJ...989..215F,Ambrosone2024JCAP...09..075A,Murase2020PhRvL.125a1101M}. The IceCube data requires $\zeta \simeq 0.1$, implying that only a fraction of AGNs are as efficient as NGC 1068 in accelerating non-thermal protons. Such a parameter is necessary to reconcile the diffuse neutrino flux with the observed neutrino point sources, since assuming all AGNs operate at the same high efficiency would exceed the diffuse background.
The physical origin of this suppression factor remains largely uncertain, but it may be related to the activity or efficiency of magnetically driven turbulence. 

The diffuse neutrino flux is primarily dominated by AGNs with X-rays luminosities in the range of $10^{42}-10^{44}\, {\rm erg \, s}^{-1}$, as this population represents the peak of the luminosity-weighted number density (i.e., the product of the AGN number density and their individual luminosities). In these sources, the $p\gamma$ interaction reaches the saturation regime as it dominates over the Bethe-Heitler process, where the reaction efficiency approaches $100\%$, at a characteristic rest-frame neutrino energy of approximately 50 TeV (corresponding to proton energy of ${\rm 1\, PeV}$). Considering that the cosmological evolution of AGN activity typically peaks at a redshift of $z \approx 0.5$ \citep{Ueda2014ApJ...786..104U}, this 50 TeV rest-frame feature is cosmologically shifted to an observed energy of $E_{\nu,  {\rm obs}} = E_{\nu} / (1+z) \approx 30\,  {\rm TeV}$, thereby naturally explaining the spectral hump observed at $\sim 30$ TeV. The predicted neutrino spectral index below 30 TeV is harder than the MESE analysis result. Note that we adopt a single-zone model for simplicity and do not include $pp$ interactions from the accretion disk embedded within the corona. Given the significantly higher proton density in the disk, the $pp$ interactions could be enhanced below $\sim$ 30 TeV. This additional component could soften the resulting neutrino spectrum below the break.
\begin{figure}
    \centering
    \includegraphics[width=0.5\textwidth]{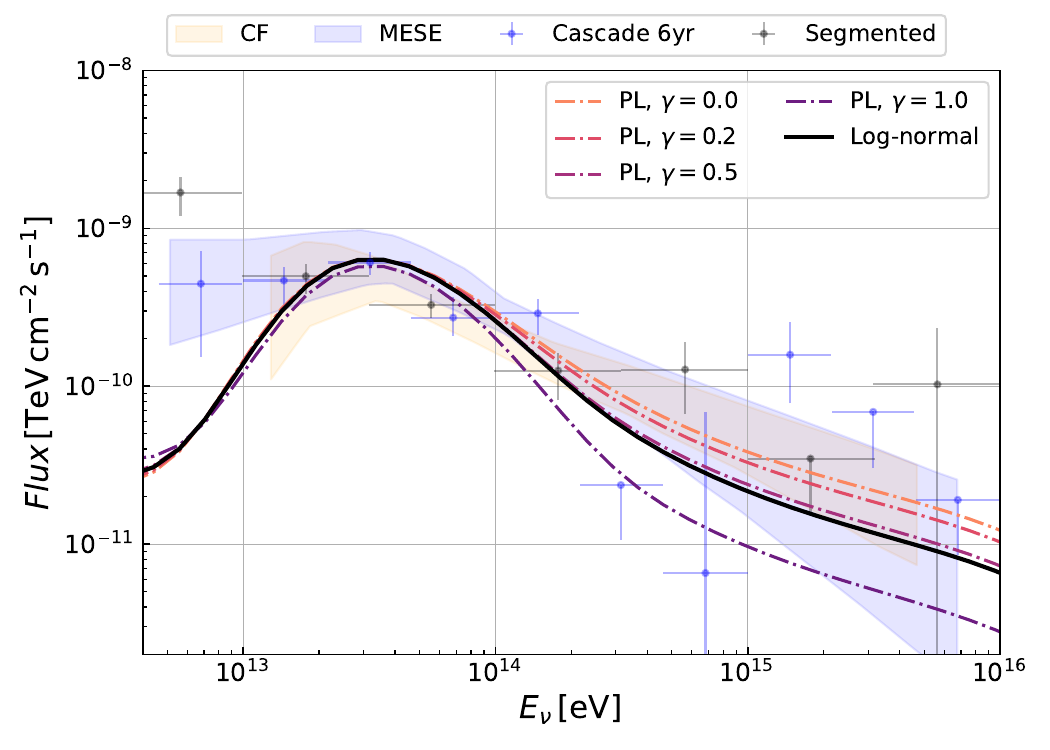}
    \caption{Diffuse all-flavor neutrino fluxes from the AGN population assuming power-law distributions for the magnetization with different power-law indices, as shown in colored dot-dashed curves. For comparison, we also plot the neutrino fluxes from the AGN coronae assuming that the magnetization distribution follows the log-normal distribution, as shown by the solid black curve. The others keep the same with Fig.\ref{Fig:Diffuse1}.} 
    \label{Fig:Diffuse2}
\end{figure}

\begin{figure}
    \centering
    \includegraphics[width=0.5\textwidth]{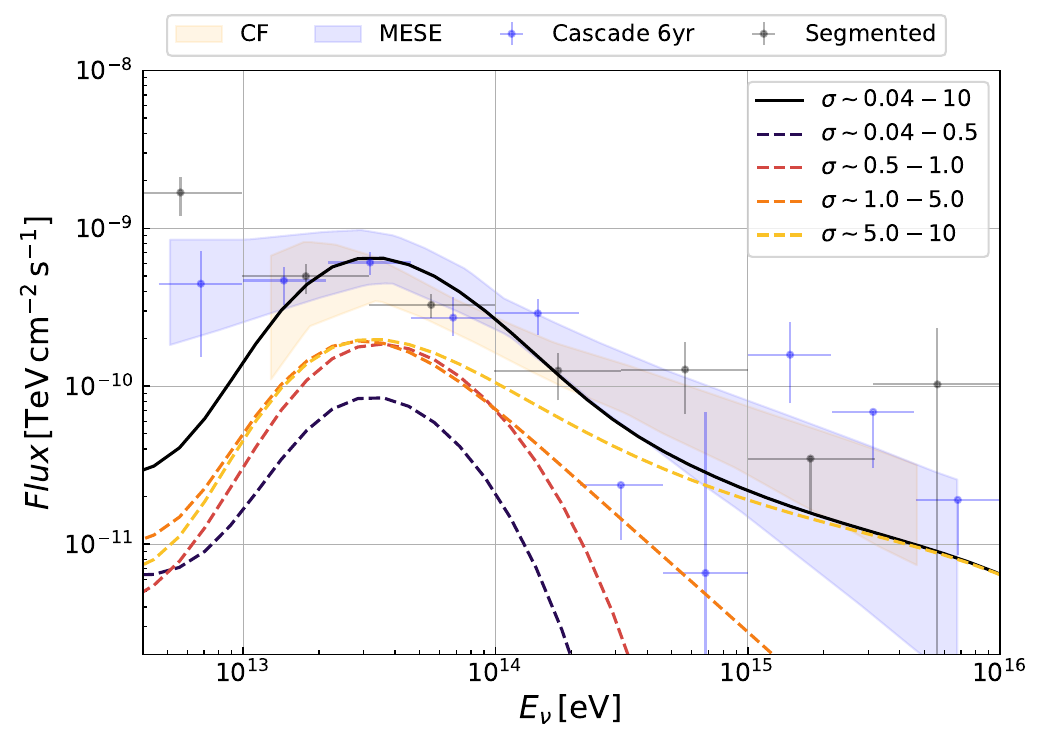}
    \caption{Same as the Fig.\ref{Fig:Diffuse1} but characterizing the contribution of each magnetization range. The colored dashed curves denote the contributions from AGNs in the different ranges of the magnetization $\sigma$, while the black solid curve represents the sum of them.} 
    \label{Fig:Diffuse3}
\end{figure}
The population of relatively low X-ray luminosity AGNs with $L_X \sim 10^{41}-10^{42}\,{\rm erg\,s^{-1}}$ constitutes the primary contribution to the diffuse neutrino flux above $\sim 100\,{\rm TeV}$, particularly for highly magnetized sources in which the maximum proton energy exceeds $\sim 10\,{\rm PeV}$ (shown by the right bottom corner in Fig.\ref{Fig:Ep_max}).   According to our model, ultra-high-energy astrophysical neutrinos are predominantly emitted by AGNs with relatively low X-ray luminosities and high magnetizations. On the other hand, we adopt a relatively broad log-normal distribution with $s \sim 1$ to ensure the presence of AGNs spanning a wide range of magnetization values. Interestingly, we also tested alternative magnetization distributions, including log-normal distributions with different central values ($10^{\mu} \sim 0.5,\,1$) and normalized power-law distributions of the form $P(\sigma)\propto\sigma^{-\gamma}$ with $0 \lesssim \gamma \lesssim 1 $. All of these tests yield neutrino spectra that are very similar to those obtained with the fiducial log-normal distribution, as long as the magnetization distribution remains sufficiently flat and broad.
Fig. \ref{Fig:Diffuse2} presents the integrated diffuse neutrino flux characterized by a power-law magnetization distribution. {We find that a hard power-law index with $\gamma \lesssim 1$ is needed to account for the diffuse neutrinos at high energies. If $s \lesssim 0.6$ in a log-normal distribution or $\gamma \gtrsim 1$ in power-law distribution, the cumulative diffuse neutrino flux will not be able to account for the neutrino signals observed by IceCube. This is due to the lack of a sufficient number of highly magnetized coronae with $\sigma \gtrsim 1$. To assess this quantitatively, by integrating the contributions from AGN coronae within $\sigma \sim 1-10$ under such distribution parameters, we find that a fraction of $\gtrsim 40\%$ of AGNs have such high magnetization.}
Fig.\ref{Fig:Diffuse3} presents the diffuse neutrino flux contributed by AGNs in different magnetization ranges, assuming that the magnetization $\sigma$ follows a log-normal distribution with $\mu = -1$ and $s = 1$. The colored dashed curves show the individual contributions from different $\sigma$ bins, while the black curve represents the total diffuse flux. It shows that the diffuse neutrinos are dominated by AGNs with $\sigma \gtrsim 0.5$.

\section{Conclusions and discussions}
\label{Sec:4}
In this paper, we investigate  whether the population of AGNs can account for the diffuse high-energy neutrinos within the framework of the turbulent coronal model. For each individual AGN, protons are accelerated by turbulence in a magnetized corona, and their acceleration properties are primarily determined by two independent parameters: the X-ray luminosity $L_X$, and the coronal magnetization $\sigma$. High-energy neutrinos are produced via $p\gamma$  interactions between relativistic protons and soft photon fields originating from the accretion disk and the corona. 
We compute the resulting diffuse neutrino spectrum by convolving the neutrino emission from individual AGN with the AGN X-ray luminosity function and an assumed distribution of the coronal magnetization across the AGN population. 
Previous studies have generally assumed a proton (ion)–electron composition for the corona in calculating the diffuse neutrinos from AGNs. It is found that such models  are insufficient to account for diffuse neutrinos above 100 TeV and some additional source classes, such as blazars, are usually invoked to explain the high-energy component. Motivated by the recent detection of {$\sim $} 100 TeV neutrinos from NGC 7469, we propose that AGN coronae with higher magnetization could account for the high-energy diffuse neutrinos over 100 TeV, allowing the AGN population to possibly explain the total diffuse neutrino flux. Note that the distribution of the magnetization of AGN coronae is largely unknown, so we test different forms for this distribution. We find that, to match the flux of diffuse neutrinos above 100 TeV,  a significant fraction of AGNs should have high magnetization, requiring that   the magnetization distribution should be sufficiently flat and broad.  

We also find that a suppression factor of $\zeta\sim 10\%$ is required for AGNs to produce a neutrino flux in agreement with the observed one. 
This implies that AGN coronae may produce neutrinos intermittently, i.e., the neutrino production efficiency should change substantially on timescale of decades, or that only a fraction ($\sim 10\%$) of AGNs can produce neutrinos.
It was recently shown that high-confidence extragalactic neutrino emitters tend not only to have higher hard X-ray fluxes but also to be more variable in mid-infrared (MIR) than other AGNs in the \textit{Swift} BAT AGN Spectroscopic Survey \citep{Zhou2025arXiv251116869Z}. The MIR variations could reflect the long-term fluctuations of the activity of AGN coronae, i.e., the AGN coronae accelerate high-energy protons only during part of the time although they still emit X-ray emission. The underlying reason is still unknown; it could be related to the intermittent activity of magnetic field reconnection/turbulence. Alternatively,  other energy dissipation mechanisms, such as dissipation of turbulence driven by magneto-rotational instability \citep{Jiang2014ApJ...784..169J},  evaporation of the inner accretion disk \citep{Meyer2000A&A...361..175M,2000A&A...360.1170R}, and/or the advection of inner accretion region may heat coronal electrons and power the X-ray coronae \citep{Merloni2000MNRAS.318L..15M,Kawabata2010PASJ...62..621K}, particularly in weakly magnetized coronae. 
In other words, coronae with weak magnetization could still produce bright X-ray emission with the absence of neutrino emission.

Interestingly,  our model can naturally explain the peak of the diffuse neutrino spectrum at $\sim 30$ TeV. The diffuse neutrino flux below 100 TeV is  dominated by AGNs with luminosities in the range of $10^{42}-10^{44}\,{\rm erg\, s^{-1}}$. For these AGNs, the $p\gamma$  reaction efficiency approaches nearly
100\% at a characteristic rest-frame neutrino energy of
approximately 50 TeV. Considering  the cosmological evolution of AGN activity, the peak shifts to $\sim 30\, {\rm TeV}$, coincident with the observation peak of the diffuse neutrino spectrum. 

\begin{acknowledgments}
We would like to thank Martin Lemoine, Mou-Yuan Sun, Kohta Murase and the anonymous referee for
valuable discussions. This work is supported by the
National Natural Science Foundation of China (grant
Nos. 12121003, 12333006 and 12393852), National SKA Program of China (2025SKA0110104) and  the
Fundamental Research Funds for the Central Universities (KG202502). We are grateful to the High Performance Computing Center of
Nanjing University for doing the numerical calculations in this
paper on its blade cluster system.
\end{acknowledgments}

\appendix
\section{Photon Field}
\label{Sec:Photon Field}
The soft photon field serving as the scattering target can be divided into two components. The first component is coronal X-ray emission, whose spectrum can be modeled by power law with an exponential cut-off. The photon index, $\Gamma_X$ , is correlated with $\lambda_{\rm Edd}$ as $\Gamma_X \approx 0.167 \times {\rm log}(\lambda_{\rm Edd}) + 2.0$ \citep{Trakhtenbrot2017MNRAS.470..800T}, and the cutoff energy is given by $\varepsilon_{X,{\rm cut}} \sim −74\, {\rm log}(\lambda_{\rm Edd} + 1.5 \times 10^2\, {\rm keV} $. $\lambda_{\rm Edd} = L_{\rm bol}/L_{\rm Edd}$, where $L_{\rm bol}$ and $L_{\rm Edd}$ are bolometric and Eddington luminosities, respectively. 

The second component is the optical–ultraviolet (OUV) photon field, which manifests as the so-called Big Blue Bump (BBB) in AGN spectra and originates from the accretion disk.  We model the OUV radiation a multi-temperature black-body emission with the maximum temperature near the central supermassive black hole $T_{\rm disk} \approx 0.49 (G M_{\rm BH} \dot{M}/(72 \pi \sigma_{\text{SB}} R_S^3))^{1/4} {\rm K}$ \citep{Pringle1981ARA&A..19..137P}. The temperature of the disk can be expressed as $T(r)\approx T_{\rm disk} (r/R_S)^{-3/4}$. Here, $M_{\rm BH}$ is the SMBH mass,  $R_S = 2GM_{\rm BH}/c^2$ is the Schwarzschild radius, and $\sigma_{\text{SB}}$ is the Stefan-Boltzmann constant. For a standard disk, one may use $\dot{M}\approx L_{\rm bol}/(\eta_{\rm rad}c^2) = \lambda_{\rm Edd}L_{\rm Edd}/(\eta_{\rm rad}c^2)$ with a radiative efficiency of $\eta_{\rm rad} = 0.1$.
We can calculate the disk luminosity as 
\begin{equation}
    L_{\rm disk} = \frac{8\pi^2 h\nu^3}{c^2}\int_{R_{\rm in}}^{R_{\rm out}}\frac{rdr}{e^{(h\nu/kT(r))}-1},
    \label{Eq:mbb}
\end{equation}
where $R_{\rm in}$ represents the inner edge of the disk, or marginally stable orbit of the disk. The size of $R_{\rm in}$ depends on the dimensionless spin parameter $a_{\star} \equiv a/R_{\rm g}$, where $a = J/cM_{\rm BH}$ ($J$ being the BH angular momentum).  When $a_{\star} = 0$, we have a non-rotating (Schwarzschild) BH, and $R_{\rm in} = 6R_{\rm g}$. The maximum value of $a_{\star}$ is less than 1 \citep{Thorne1974ApJ...191..507T}, but when $a_{\star} \simeq 1$, we have
a maximally rotating BH, and $R_{\rm in} \simeq R_{\rm g}$. In this paper, we utilize $R_{\rm in} \simeq 2R_{\rm g}$ as a fiducial value of inner disk radius \citep{Zimmerman2005ApJ...618..832Z}. Observations have revealed the relationship between the X-ray luminosity $L_X$ and $L_{\rm bol}$ \citep{Hopkins2007ApJ...654..731H}, by which the disk-corona SEDs can be modeled as a function of $L_X$.

\section{Cooling and escape processes of protons in the coronae} 
\label{Sec:Loss}

For the cooling processes of cosmic ray protons, we consider inelastic collisions ($pp$), photomeson production ($p\gamma$), Bethe-Heitler pair production (B-H), and proton synchrotron radiation. For escape terms, we consider diffusion and free-fall (infall to the BH) as the primary escape processes.
Firstly, we consider the $p\gamma$ process and Bethe–Heitler pair production. The soft photon field has been discussed in previous Appendix.\ref{Sec:Photon Field} Thus, the timescale of $p\gamma$ is  calculated by
\begin{equation}
      t_{p\gamma}\approx \left(n_{\gamma}\hat{\sigma}_{p\gamma}\kappa_{p\gamma}c\right)^{-1}, 
    \label{Eq:timescale pgamma calculated} 
\end{equation}
where $\hat{\sigma}_{p\gamma} \approx 5\times 10^{-28}~\rm{cm}^{2}$ is the cross section for the photomeson process, $\kappa_{p\gamma} \sim 0.2$ is the inelasticity for $p\gamma$. The Bethe-Heitler timescale $t_{\rm B-H}$  is
\begin{equation}
\begin{aligned}
    t_{\rm B-H}\approx (n_{\gamma}\hat{\sigma}_{\rm B-H}c)^{-1},
    \label{Eq:timescale BH}
\end{aligned}    
\end{equation} 
where $\hat{\sigma}_{\rm B-H}\approx 0.8\times10^{-30} \rm{cm}^{2}$ is the effective cross section for the Bethe-Heitler process. $n_{\gamma} = L_{\varepsilon_{\gamma}}/4\pi R^2c\varepsilon_{\gamma} $ is the number density of the target photon field with $L_{\varepsilon_\gamma}$ the luminosity at photon energy $\varepsilon_\gamma$.  

Then we try to determine the timescale of the $pp$ process. The timescale of $pp$ collision is shown by

\begin{equation}
\begin{aligned}
    t_{pp} \approx (n_{p}\hat{\sigma}_{pp}\kappa_{pp}c)^{-1},
\label{Eq:timescale pp}
\end{aligned}
\end{equation}
where $\hat{\sigma}_{pp} \simeq 4\times 10^{-26} \rm{cm}^{2}$ and $\kappa_{pp} \approx 0.5$ are cross section and inelasticity for $pp$ process, respectively \citep{Kelner2006PhRvD..74c4018K}.
The proton synchrotron timescale is
\begin{equation}
    t_{p,\rm{syn}} = \frac{6 \pi m_p c}{\gamma_p \sigma_T B^2} \left( \frac{m_p}{m_e} \right)^2.
\label{Eq:timescale_syn}
\end{equation}

The timescales of in-fall is $t_{\rm fall} \approx R/V_{\rm R}$. $V_{\rm R} \simeq \alpha V_k/2$ is the radial velocity where $V_k$ is Keplerian velocity and $\alpha\simeq 0.3$ is viscous parameter. In the turbulence scenario,
particle transport occurs via scattering on magnetic inhomogeneities (mean free path $\lambda_{\rm scatt}$) and turbulent diffusion ($\kappa_{\rm turb} \sim \ell_cv_A/3$). Since $\lambda_{\rm scatt} \sim r_L^{1/3}\ell_c^{2/3}$, escape is mainly governed by turbulence. We therefore adopt $\kappa = \kappa_{\rm turb}  + \lambda_{\rm scatt}c/3$,  giving an diffusion escape timescale $t_{\rm diff} = R_{\rm c}^2/(2\kappa)$.

The total cooling timescale is $t_{\rm cool}^{-1} = t_{p,{\rm syn}}^{-1}+t_{pp}^{-1}+t_{\rm B-H}^{-1}+t_{p\gamma}^{-1}$  and the total escape timescale is calculated as  $  t_{\rm esc}^{-1} = t_{\rm diff}^{-1}+ t_{\rm fall}^{-1}$. Hence, we can calculate the timescale of the total energy loss as
\begin{equation}
    t_{\rm loss}^{-1} = t_{\rm cool}^{-1}+ t_{\rm esc}^{-1}.
\label{Eq:loss}
\end{equation}

\bibliography{main.bib}{}
\bibliographystyle{aasjournalv7}
\end{document}